\documentclass[12pt,eqsecnum,nofootinbib,floats,article,aps,prd,floatfix,titlepage,superscriptaddress]{revtex4} 
\usepackage{epsfig}
\usepackage{graphics}
\usepackage{bm}
\usepackage{amsmath}

\begin{document}

\title{Tumbling through a landscape: Evidence of instabilities in
  high-dimensional moduli spaces}

\author{Brian Greene}
\email{greene@phys.columbia.edu}
\affiliation{Physics Department, Columbia University, New York, New York 10027, USA}
\author{David Kagan}
\email{dkagan@umassd.edu}
\affiliation{Department of Physics,
University of Massachusetts Dartmouth, \\ North Dartmouth MA 02747, USA}
\author{Ali Masoumi}
\email{ali@phys.columbia.edu}
\affiliation{Physics Department, Columbia University, New York, New York 10027, USA}
\author{Dhagash~Mehta}
\email{dbmehta@syr.edu}
\affiliation{Department of Physics, Syracuse University, Syracuse, NY 13244, USA}
\author{Erick J. Weinberg}
\email{ejw@phys.columbia.edu}
\affiliation{Physics Department, Columbia University, New York, New York 10027, USA}
\author{Xiao Xiao}
\email{xx2146@columbia.edu}
\affiliation{Physics Department, Columbia University, New York, New York 10027, USA}

\begin{abstract}

We argue that a generic instability afflicts vacua that arise in
theories whose moduli space has large dimension. Specifically, by
studying theories with multiple scalar fields we provide numerical
evidence that for a generic local minimum of the potential the usual
semiclassical bubble nucleation rate, $\Gamma = A \,e^{-B}$, increases
rapidly as function of the number of fields in the theory.  As a
consequence, the fraction of vacua with tunneling rates low enough to
maintain metastability appears to fall exponentially as a function of the
moduli space dimension.  We discuss possible implications for the
landscape of string theory. Notably, if our results prove applicable
to string theory, the landscape of metastable vacua may not contain
sufficient diversity to offer a natural explanation of dark energy.

\end{abstract}

\maketitle
\section{Introduction}
The discovery that string theory admits an enormous number of flux
vacua \cite{Strominger:1986uh,Polchinski:1995,Giddings:2001yu} has
played a formative role in the theory's development for well over a
decade. Aspects of these vacua, from their phenomenological and
cosmological properties to their distribution and statistical
features, have been extensively studied
\cite{Grana:2005jc,Douglas:2006es}.  Yet, due to the complexity of
this landscape of vacua, many basic questions remain. In this paper we
undertake a study, pursued from a variety of perspectives in a number
of works \cite{Kachru:2003sx,Ceresole:2006iq,Dine:2007er,
  Sarangi:2007jb, Tye:2007ja, Podolsky:2008du, Brown:2010bc,
  Brown:2007zzh},
that is of potential
relevance to one such vital question: Do we expect these vacua to be
long-lived?  As a direct analysis would present formidable challenges,
we instead consider generic, field theoretic models of the landscape
and study how the stability of vacua varies as the dimension of the
moduli space (the number of fields) increases. Our results suggest
that tunneling rates, and hence vacuum instability, grow so rapidly
with the number of moduli that the probability of a given local
minimum being metastable is exponentially small.

In field theory, vacuum decay by quantum tunneling was studied by
Coleman~\cite{Coleman:1977py,Callan:1977pt}.  He showed that the decay
proceeded by the nucleation of bubbles of a lower vacuum inside the
original false vacuum.  In the semiclassical approximation the
nucleation rate per unit volume is governed by a bounce solution of
the Euclidean field equations.  It can be written in the form $\Gamma=
A e^{-B}$, where $A$ depends on the determinant of fluctuations around
the bounce solution and $B$ is the Euclidean action of the bounce.
The analysis was extended to include gravitational effects by Coleman
and De Luccia~\cite{Coleman:1980aw}.  For decay from a de Sitter vacuum the
resulting corrections to the nucleation rate are typically small
unless the potentials are Planckian in scale or the bubbles nucleate
with a size comparable to the horizon length.  With unusually flat
potential barriers it can happen that there is no Coleman-De Luccia
bounce, but in such cases there is always a
Hawking-Moss~\cite{Hawking:1981fz} solution corresponding to a process
in which an entire horizon volume fluctuates to the top of the
potential barrier.

In string theory the large number of vacua and moduli fields
complicates the situation, but at the same time opens up lines of
attack based on statistical analysis. For example, Denef and Douglas
\cite{Denef:2004ze} proposed a method of calculating the density of
 flux vacua in the string landscape in terms of the
K\"ahler potential on the moduli space of a given Calabi-Yau
compactification. Their work showed that a sharp accumulation of vacua
generally occurs near the conifold locus in moduli space.  Dine et al. 
\cite{Dine:2007er} used scaling arguments to conclude that vacua in the string 
landscape with small cosmological constant become 
unstable when fluxes are large relative to their compactification volume.
Chen et al.~\cite{Chen:2011ac} and Marsh et al.~\cite{Marsh:2011aa} showed
that increasing the number of moduli fields suppresses exponentially
the chance that a randomly chosen critical point will be a minima,
with the suppression growing for vacua with high energy.

In this paper we also pursue a statistical approach, focusing our
analysis on effective field theory models of many-dimensional moduli
spaces. We provide numerical evidence that the rate for tunneling out
of a typical false vacuum grows rapidly as a function of the number of
moduli fields.  Specifically, the fraction of vacua with tunneling
rates low enough to maintain metastability appears to fall as an
exponential of a power of the moduli space dimension.

In Sec.~\ref{many-fields} we describe our approach for estimating
tunneling rates in field theoretical models of high-dimensional moduli
spaces. Our numerical methods and results are described in
Sec.~\ref{numerics}. These results reveal a general feature of
high-dimensional field theories and are independent of applications to
the string landscape. In Sec.~\ref{implications} we discuss the
efficacy of using random potentials to model the string landscape, and
emphasize various considerations that would need to be resolved before
a direct application could be justified. In
Sec.~\ref{Multiverse} we estimate the maximum dimension of moduli
spaces whose associated flux landscape would be expected to
generically contain metastable vacua, assuming that our numerics and
extrapolations are applicable.  The result is a drastic reduction in
the number of metastable vacua.  Finally, in Sec.~\ref{Conclusions} we
suggest future directions for studying these models and summarize
our conclusions.

\section{Vacuum Decay}
\label{many-fields}

We consider the dynamics of a moduli space spanned by $N$ scalar
fields $\phi_j$ with a Lagrangian
\begin{equation}
  {\cal L} = \frac12 \sum_{j=1}^N \, \partial_\mu \phi_j \partial^\mu \phi_j    
        - V(\phi_1,\phi_2,\dots, \phi_N)   \, .
\end{equation}
The potential $V$ will in general have many local minima that
correspond to metastable false vacua.  Let us consider one of these
which, by a shifting of the field variables, can be taken to lie at
the origin of field space, $\phi=0$.  Assuming the potential to be
smooth at this point, we can expand it in a power series
\begin{equation}
    V = \lambda \left( \sum_i  A_{ii}^{(2)} \phi_i^2 v^2 
      + \sum_{ijk} A_{ijk}^{(3)} \phi_i \phi_j \phi_k v
      + \sum_{ijkl} A_{ijkl}^{(4)} \phi_i \phi_j \phi_k \phi_l 
     + ... \right)  \, .
\label{Vexpansion}
\end{equation}
Here $v$, with dimensions of mass, is a characteristic distance in
field space that corresponds to a typical distance between stationary
points of $V$.  We have also extracted a dimensionless constant
$\lambda$, to be chosen so that the dimensionless coefficients of the
power series in brackets are of order unity.  Finally, we have used
the freedom to make an O($N$) transformation on the fields to
eliminate off-diagonal terms in the quadratic term of the power
series.

The exponent $B$ in the bubble nucleation rate is the action of the
Euclidean bounce solution.  We ignore gravitational effects and assume
O(4) symmetry, with the fields being functions only of $s=\sqrt{{\bf
    x}^2 + x_4^2}$.  The bounce then satisfies
\begin{equation}
    {d^2\phi_j \over ds^2}   + \frac{3}{s}\,  {d\phi_j \over ds} 
   = {\partial V \over \partial \phi_j} \, .
\end{equation}
The boundary conditions are that $\phi(\infty)=0$, its false vacuum
value, and that $\phi'(0)=0$.  The actual value of the field at the
origin is not determined in advance, but must be a point on the
opposite side of the potential barrier from the false vacuum.  Note
that, except in the thin-wall limit, $\phi(0)$ is never equal to the
true vacuum value.

We will consider large ensembles of potentials, with the coefficients
in the power series chosen randomly, as described in more
detail in the next section.  Ideally, we would calculate the nucleation 
rate for each potential by solving the bounce equations.  
However, finding bounce solutions in
a theory with more than one scalar field is a daunting numerical
problem.  Doing so for a large sample of potentials is clearly
infeasible.  Instead, we invoke a more easily calculable proxy that
can provide an indication of how the decay rate varies with the
number of fields.

A bounce solution may be viewed as containing a wall region, in
which the fields pass through the barrier in $V(\phi)$, and an
interior region where the fields are close to their values in the new
vacuum.  The distinction between the two regions becomes exact in the
thin-wall approximation, which is valid in the limit where $\epsilon$,
the difference between the energy densities of the true and false
vacua, tends to zero.  In this approximation the fields in the
interior region take on exactly their true vacuum values.  The radius
$R$ of the bounce is determined by a balance between the negative
action in the bounce interior and the positive action in the wall.
With a single scalar field, this wall action is the product of 
the three-dimensional area of bounce and a surface tension that is well 
approximated by 
\begin{equation}
    \sigma = \left|\int_{\phi_{\rm fv}}^{\phi_*} d\phi 
     \, \sqrt{2[V(\phi) - V_{\rm fv}]} \right| \, ,
\label{sigma-def}
\end{equation}
where $\phi_*$ denotes the point on the true vacuum side of the
barrier such that $V(\phi_*) = V_{\rm fv}$.
With additional scalar
fields, $\sigma$ is obtained by integrating along a path in field
space running from the false vacuum to a point near the true vacuum,
with the path and endpoint chosen to minimize the integral.

The net result is that $R = 3\sigma/\epsilon$, while the tunneling
exponent is
\begin{equation}
      B = {\pi^2\over 2} \sigma R^3 
   = {27\pi^2 \sigma^4 \over 2 \epsilon^3} \, .
\label{twaB}
\end{equation}
Thus, in the rare cases in which we have two almost degenerate vacua, the
thin-wall approximation is applicable, $B$ is large, bubble nucleation
is greatly suppressed, and the false vacuum is long-lived.  Since our
focus is on effects that enhance bubble nucleation, this approximation
is not of direct interest to us.  Nevertheless, we can draw some
useful insight from it.  Equation~(\ref{twaB}) shows a strong dependence
on $\epsilon$.  This cannot continue outside
the thin-wall approximation, because then the field in the bounce
solution never reaches the true vacuum, and so cannot be directly
sensitive to the value of $\epsilon$.  On the other hand, the surface
tension is closely related to the form of the potential barrier and
should continue to be relevant.

Outside the thin-wall limit, the boundary between the wall and the
bounce interior is not well defined.  A reasonable prescription would
be to take it to be the hypersurface in Euclidean space on which
$V(\phi)$ is equal to its false vacuum value.  We could then define a
quantity $\sigma$ as before, with the integration path running from
the false vacuum to a point on the hypersurface $\Sigma$ in field
space, lying on the other side of the potential barrier, on
which $V(\phi)=V_{\rm fv}$.  The path and the specific endpoint on
$\Sigma$ would be chosen to minimize the integral.

Unfortunately, performing the required minimization for a large
ensemble of potentials is still calculationally infeasible.  However,
a plausible approximation is at hand.  We might expect the minimizing
path to pass though the region in field space where the barrier in
$V(\phi)$ is lowest.  This suggests considering a straight-line path
running from $\phi=0$, though each saddle point, $\phi_{\rm
  sp}$, on the surrounding barrier, and ending on $\Sigma$.  In fact, since on
average the contributions from the segments before and after the
saddle point will be equal,  we follow a slightly simpler
approach.  We integrate over
only the first part of the path, extract factors of $\lambda$ and
$v$, and define the line integral
\begin{eqnarray}
     \tilde\sigma &=& 2 \int_P 
        d\phi\, \sqrt{2[V(\phi) - V(0)]}  \cr
       & \equiv& \sqrt{\lambda} \, v^3\, \tilde s  \, ,
\end{eqnarray}
where $P$ is a straight-line path in field space running from 
the false vacuum at $\phi=0$ to the saddle point.
There may be many saddle points, and thus many such paths.  We expect
the tunneling rate to be controlled by the path with the lowest
tunneling integral, and define the corresponding value of $\tilde s$
to be
\begin{equation}
       \tilde s_{\rm min} = s  \, .
\end{equation}

Standard scaling arguments show that $B$ is inversely
proportional to $\lambda$, but independent of $v$, although $v$ does
affect the prefactor $A$ in $\Gamma$.  
Outside the thin-wall limit, the typical bounce radius is $R\sim 
k (\sqrt{\lambda}\,v )^{-1}$, where $k$ is a numerical factor of order
unity. This suggests that 
\begin{equation}
     B \sim \pi^2 R^3 \tilde \sigma 
          \sim {\pi^2 \over \lambda} \,k^3 \, s  \, ,
\label{B-estimate}
\end{equation}
where we have included a factor of $\pi^2$ because of the
four-dimensional spherical symmetry.  A simple test of this estimate
can be obtained by numerically evaluating $B$ and $s$ for a
single-field potential with cubic and quartic terms.  Inserting the
results into Eq.~(\ref{B-estimate}) and taking $v$ to be half the
difference between the true and false vacuum values of $\phi$, one 
obtains values of $k$ that vary between 5.2 and 6.0 over a wide
range of parameters away from the thin-wall limit.  This suggests 
\begin{equation}
    B \sim 10^3 \, {s \over \lambda}  \, .
\label{estimateB}
\end{equation}

The dilute-gas approximation that underlies the semiclassical approach
to bubble nucleation breaks down when $B$ is comparable to or less than 
unity.  In this regime the metastability of the false vacuum has essentially 
disappeared.  The numerical studies that we describe in the next
section show that when many scalar fields are present, the overwhelming
majority of potentials lead to a value of $B$ too small to maintain 
metastability with any plausible value of $\lambda$.

\section{Numerical Studies}
\label{numerics}

We numerically studied ensembles of theories with potentials of the
form of Eq.~(\ref{Vexpansion}), using $s$ as an indicator of the 
vacuum stability.  To make the calculations manageable we truncated
the power series at the quartic
terms.
An ensemble was defined by taking the $A^{(n)}$ to be 
random numbers uniformly distributed over ranges defined by 
\begin{eqnarray}
        A_{ii}^{(2)} & \in &  [0,a_2]   \, ,\cr
        A_{ijk}^{(3)}  & \in &  [-a_3,a_3]   \, ,\cr
        A_{ijkl}^{(4)}  & \in &  [-a_4,a_4] \, .
\end{eqnarray}
Allowing the $A_{ijkl}^{(4)}$ to be negative means that the truncated
potential is not necessarily bounded from below.  This is not a
concern for us, since we are only concerned with the behavior near the
minimum; at larger distances higher-order terms can provide a lower
bound on $V$.

Because we will be comparing theories with different numbers of
fields, an important issue is how to vary the $a_n$ as the number of
fields is varied.  To determine this, let us require that the typical
variation of the potential in a ball of radius $\phi_R$ in field space
be independent of $N$, the number of fields. This ensures that, for
any number $N_0 \le N$, we can recover results for $N - N_0$ fields by
considering an $N_0$-dimensional cross section of the analysis for $N$
fields. In turn, this ensures that dependencies we find on $N$ are not
due to peculiar $N$-dependent normalizations in $V$.

With this assumption, the typical value for each of the $\phi_j$ in
such a ball of radius $\phi_R$ is of order $\phi_R/\sqrt{N}$.  The
quadratic term is a sum of positive contributions and will be
independent of $N$ if $a_2$ is.  There are $N^3$ cubic terms, each of
magnitude $\phi_R^3/N^{3/2}$.  Because these can be of either sign,
they will tend to cancel, so that the effective number of terms is of
order $N^{3/2}$ and we are led to take $a_3$ to also be
$N$-independent.  A similar argument shows that $a_4$ should also be
independent of $N$.

As for the actual values of the $a_n$, note that a change of any two
of these can be absorbed by a redefinition of $\lambda$ and $v$.
Hence, there is no loss of generality in taking $a_2=a_3=1$.  We did
so, and also set $a_4=1$; the effect of other choices for $a_4$ is
described below.

\begin{figure}
   \centering
   \includegraphics[width=3.2in]{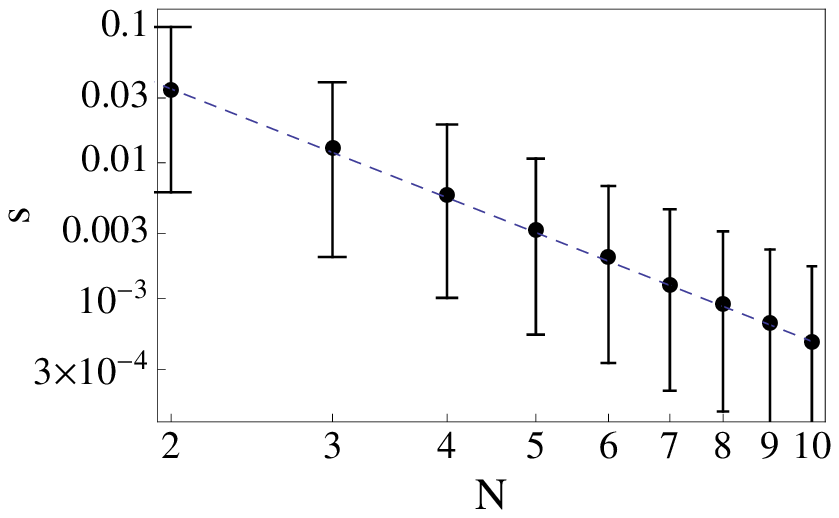}
   \includegraphics[width=3.2in]{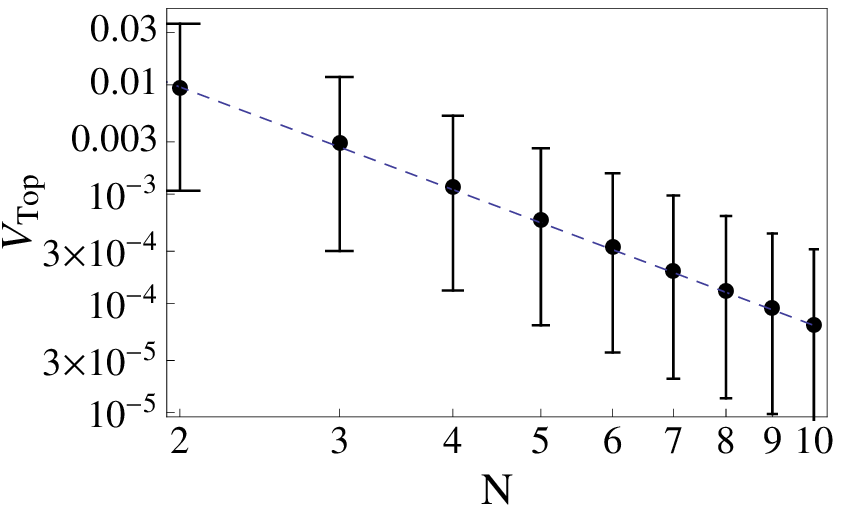}
   \caption{Left, the median value of $s$ for quartic
     potentials. Right, the median value of the lowest saddle point,
 in units of $\lambda v^4$,
     for quartic potentials.  In both cases the bars indicate the
     range from the 25th to the 75th percentile.}
   \label{QuarticTensionAndHeight}
\end{figure}

\begin{figure}
   \centering
   \includegraphics[width=3.2in]{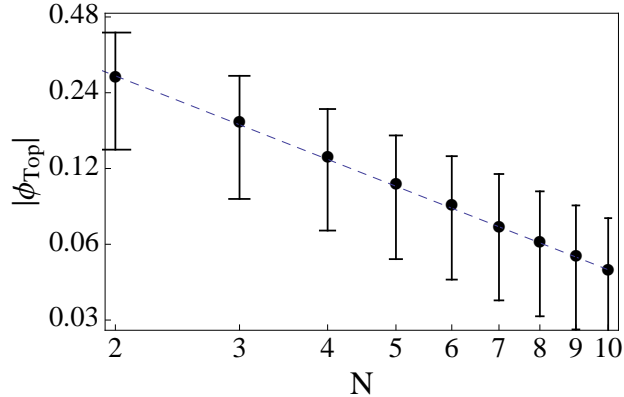}
   \caption{The median distance, in units of $v$, to the saddle point with $\tilde
     s_{\rm min}=s$ for quartic potentials.  Again, the bars indicate
     the range from the 25th to the 75th percentile.}
   \label{QuarticDistance}
\end{figure}

For a given value of $N$ we chose an ensemble of 10,000
potentials.
For each potential we found all of the stationary points\footnote{For 
a description of the numerical method for finding all such points, see
\cite{Mehta:2011xs}.  Further details will be published elsewhere~\cite{mehta-unpub}.}
and picked out the saddle points with a single negative mode. From
among the saddle points for each potential 
we found 
the one that
gave the smallest value of $\tilde s$ (i.e., $\tilde s_{\rm min} \equiv s$)
and the one that gave the lowest barrier height.
Figure~\ref{QuarticTensionAndHeight}
shows the median values of these quantities within each ensemble as
functions of $N$ for the quartic potential.  
For both quantities a
sharp decrease with increasing $N$ is clearly evident.  Not
only do the height and surface tension of these saddle points
decrease, but also they get closer to the false vacuum minimum at the
origin.  This can be seen in Fig.~\ref{QuarticDistance}, where we have
plotted the median distance to the saddle point with lowest $\tilde s$.  (A
plot of median distance to the lowest saddle point is virtually
indistinguishable.)  
All of these plots show a
power law dependence on $N$ with, e.g., 
\begin{equation}
     s_{\rm median} \approx C_{\rm tension} \, N^{-\alpha_{\rm tension}} \, .
\label{s-fit}
\end{equation}
The best fit values for the various $\alpha$ and $C$ are
shown in Table~\ref{expTable}.

\begin{table}[htdp]
\begin{center}
\begin{tabular}{|l|c|c|c|c|} \hline
                                &       ~$\alpha_{\rm tension}$~ &
  ~$\alpha_{\rm height}$~& ~$\alpha_{\rm distance}$~
    \\ \hline
        Cubic potentials     &        2.73   &   3.16       &  1.15  \\ \hline
        Quartic potentials   &        2.66   &   3.12  &  1.10    \\ \hline
        SUSY                 &        3.16   &   3.99  &  1.19    \\ \hline
\end{tabular}
\end{center}

\begin{center}
\begin{tabular}{|l|c|c|c|c|} \hline
                &       ~$C_{\rm tension}$~ & ~$C_{\rm height}$~ &
   ~ $C_{\rm distance}$~  \\ \hline
        Cubic potentials     & 0.26  &  0.090    & 0.67      \\ \hline
        Quartic potentials   & 0.22   & 0.083    & 0.60    \\ \hline
        SUSY                 & 0.25   & 0.11   &  0.60 \\ \hline
\end{tabular}
\end{center}
\caption{Best fit parameters, defined as in Eq.~(\ref{s-fit}), for
power law fits to the data in Figs.~\ref{QuarticTensionAndHeight}, \ref{QuarticDistance},
\ref{CubicTensionAndHeight}, and \ref{CubicDistance},
as well as to the data for the supersymmetric 
potentials discussed in Sec.~\ref{implications}.}
\label{expTable}
\end{table}

\begin{figure}
   \centering
   \includegraphics[width=3.2in]{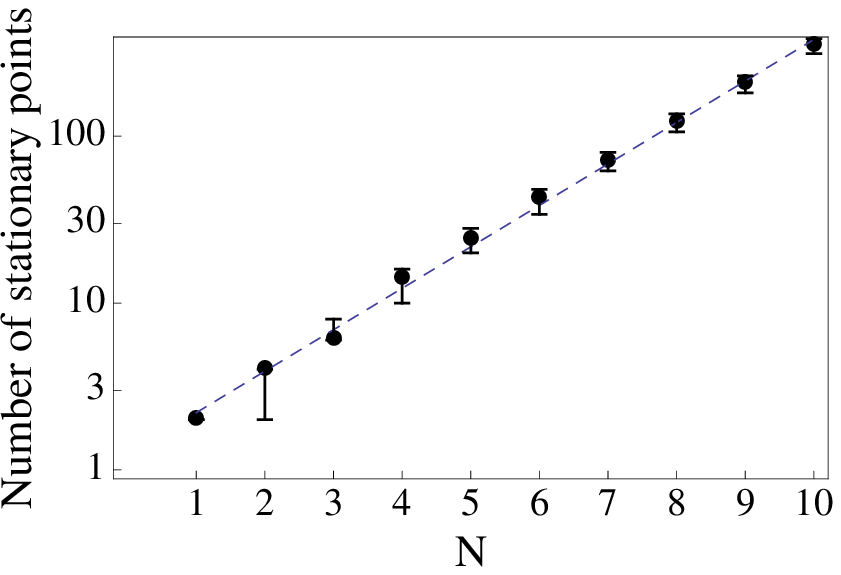}
   \caption{The median number of saddle points for quartic
     potentials, with the bars indicating the range from the 25th to
     the 75th percentile.}
   \label{QuarticExtremaMedians}
\end{figure}

We have also plotted the number of stationary points around the false vacuum, in
Fig.~\ref{QuarticExtremaMedians}.  These do not follow a power law, but are instead
closer to an exponential behavior.

\begin{figure}
   \centering
   \includegraphics[width=3.2in]{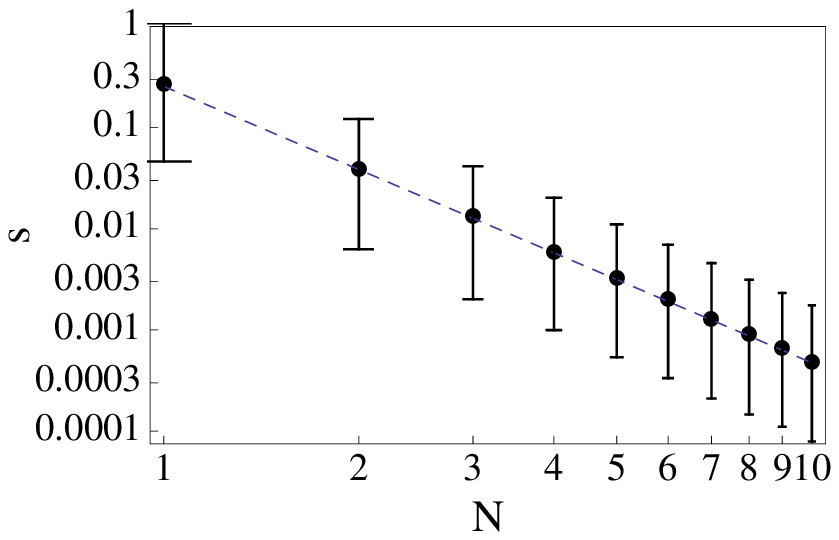}
   \includegraphics[width=3.2in]{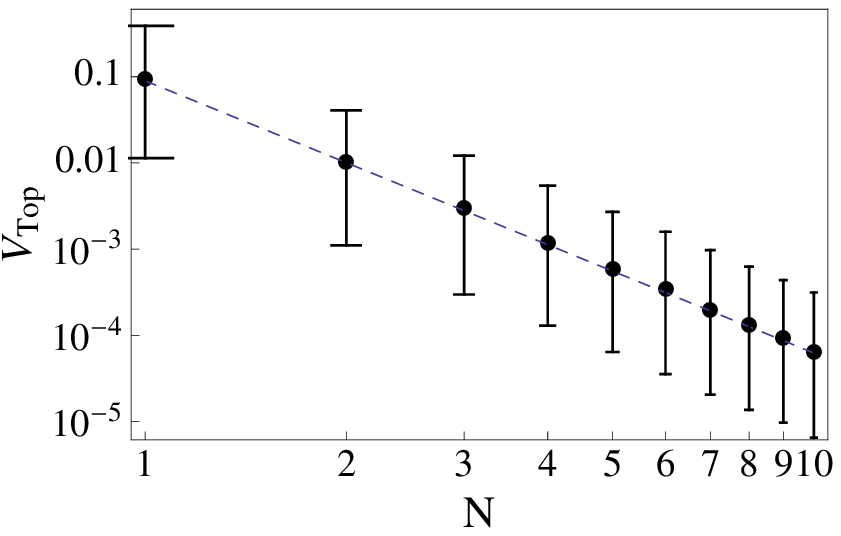}
   \caption{The same as in Fig.~\ref{QuarticTensionAndHeight}, but for
     cubic potentials.}
   \label{CubicTensionAndHeight}
\end{figure}

\begin{figure}
   \centering
   \includegraphics[width=3.2in]{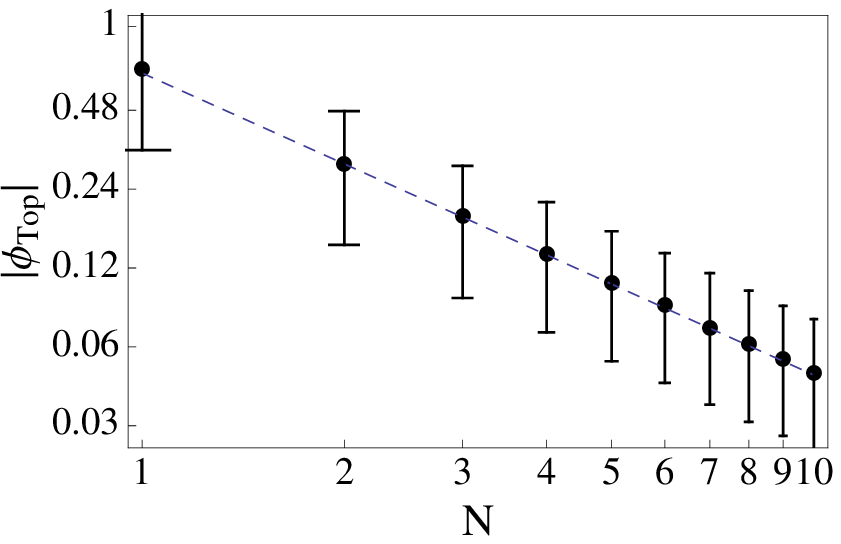}
   \caption{The same as in Fig.~\ref{QuarticDistance}, but for cubic
     potentials.}
   \label{CubicDistance}
\end{figure}

\begin{figure}
   \centering
   \includegraphics[width=3.2in]{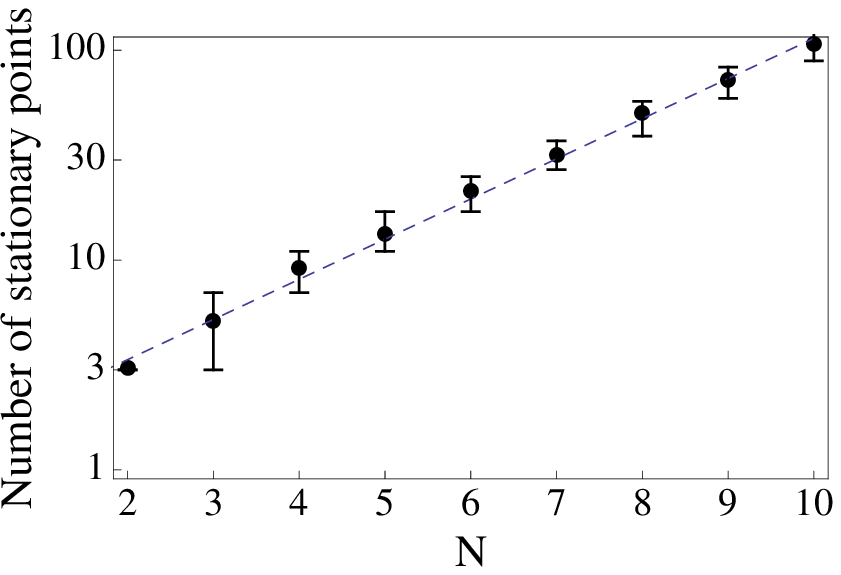}
   \caption{The same as in Fig.~\ref{QuarticExtremaMedians}, but for
     cubic potentials.}
   \label{CubicExtremaMedians}
\end{figure}

Our decision to arbitrarily terminate the expansion of the potential
with the quartic terms was motivated by considerations of
calculational practicality.  As a test of this choice, we also carried
out the calculations without the quartic terms in the potential,
retaining only the quadratic and cubic terms.  As can be seen from the
plots in Figs.~\ref{CubicTensionAndHeight}-\ref{CubicExtremaMedians}.
and the data in Table~\ref{expTable}, the results are quite similar to
those with the quartic term included.  This leads us to conclude that 
our omission of quintic and higher terms has little effect on our results.

\begin{figure}
   \centering
   \includegraphics[width=3.2in]{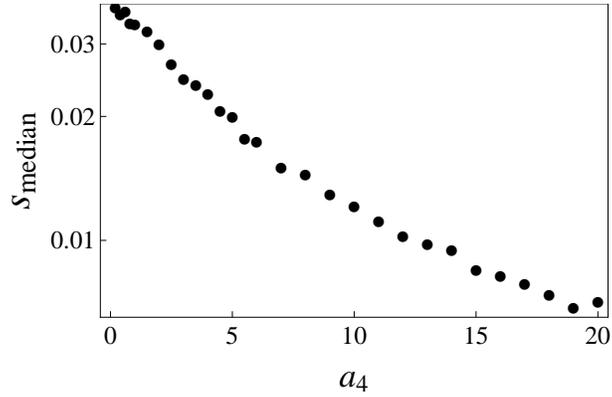}
   \caption{The dependence of the median value of $s$ on the range
     parameter $a_4$. The data shown are for $N=2$.}
   \label{vary-a4}
\end{figure}

We now return to the question of the dependence of the results on the
ranges chosen for the coefficients in the potential.  As noted
previously, any change in the constants defining the ranges of the
quadratic and cubic terms can be compensated by a rescaling of
$\lambda$ and $v$.  The dependence on the quartic coefficient range is
illustrated in Fig.~\ref{vary-a4}, where we plot the median value of
$s$ as a function of $a_4$ with $N=2$.  We see that increasing $a_4$
produces a roughly exponential decrease in the median value of $s$;
decreasing $a_4$ to zero simply reduces to the purely cubic potential.
Hence the net effect of increasing the range for $a_4$ would be to
lead to a lower tunneling exponent, thus strengthening the effects
that we find.  The results for the other quantities and other values
of $N$ give similar results.

\begin{figure}
   \centering
   \includegraphics[width=5.5in]{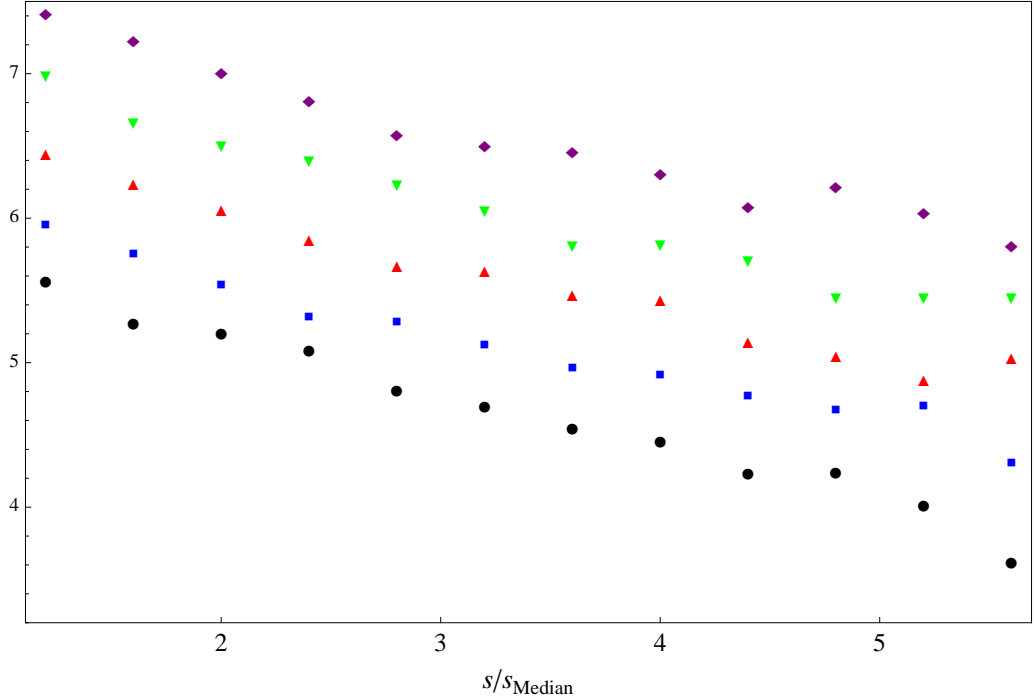}
   \caption{The distribution of values of $s$ in the ensembles for a
     quartic potential with various values of $N$.  The vertical axis
     represents the natural logarithm of the number of values in each
     bin.  For the sake of clarity, the data for different values of
     $N$ have been offset by constants, so only the slopes are
     meaningful.  Purple diamonds correspond to $N=2$; green down-pointing
     triangles, $N=4$; red up-pointing triangles, $N=6$; blue squares, $N=8$;
     and black circles, $N=10$.}
   \label{QuarticDistributions}
\end{figure}

Our data indicate that the median value of $s$ falls
rapidly as the number of fields is increased.  From this we can
conclude that at large $N$ the typical local minimum is less likely to
have a low nucleation rate and a long lifetime.  This by itself does
not tell us about the number of outliers, vacua with high values of
$s$.  To study this we examined the distributions of the
quantities that we have plotted within a given ensemble.  For example,
in Fig.~\ref{QuarticDistributions} we show the distributions of values
of $s$ for several choices of $N$ with a quartic potential.
These roughly coincide when they are plotted as functions of
$s/s_{\rm median}$.  The data suggest that the
frequency of large $s$ has an approximately exponential
falloff that can be described by 
\begin{equation}
      n(s) \approx n_0 
     \exp\left(-\gamma s/s_{\rm median}\right)  \, .
\end{equation}
The values of $\gamma$ for various values of $N$ are 
shown in Table~\ref{distribTable}.  
Similar results are found for the data on the height of and distance
to the lowest saddle point.

\begin{table}[htdp]
\begin{center}
\begin{tabular}{|c|c|c|c|} \hline
    ~$N$~    &      ~ Cubic~ &  ~Quartic~ & ~SUSY~  \\ \hline
        1       &       0.58   &  0.46  & 0.45 \\ \hline
        2       &       0.38   &  0.39  & 0.33\\ \hline
        3       &       0.40   &  0.35  & 0.33 \\ \hline
        4       &       0.34   &  0.33  & 0.32  \\ \hline
        5       &       0.37   &  0.35  & 0.32  \\ \hline
        6       &       0.37   &  0.34  &  \\ \hline
        7       &       0.38   &  0.34  &  \\ \hline
        8       &       0.38   &  0.35  &   \\ \hline
        9       &       0.38   &  0.37  &   \\ \hline
        10       &      0.35   &  0.34  &   \\ \hline
\end{tabular}
\end{center}
\caption{Best fit values for the parameter $\gamma$ in the
  distributions of surface tension $s$ for cubic and quartic non-supersymmetric
  and quartic supersymmetric potentials with various $N$.}
\label{distribTable}
\end{table}

Inserting the fit of Eq.~(\ref{s-fit}) for $s_{\rm median}$ into this
expression gives 
\begin{equation}
    n(s) \approx n_0 \exp\left(- {\gamma\over C_{\rm tension}}N^\alpha s \right) \, ,
\end{equation}     
where for convenience we have defined $\alpha \equiv \alpha_{\rm tension}$.
Extrapolating to larger $N$ and using the estimate in
Eq.~(\ref{estimateB}) suggests that the fraction of potentials with a
tunneling exponent greater than some value $\hat B$ is roughly
\begin{equation}
    f(\hat B) \sim \exp(- \beta \,N^\alpha \, \hat B)   \, ,
\label{fractionEst}
\end{equation}
where
\begin{equation}
      \beta  =  10^{-3} \, {\gamma \lambda \over C_{\rm tension}}  \, .
\end{equation}
Our numerical results suggest that $\gamma/C_{\rm tension}$ is close to unity.  Because
$\lambda$ is extracted from the shape of the landscape, there is no reason 
to expect it to be a small coupling constant.  It could well be of order 
unity, in which case $\beta \sim 10^{-3}$.
Equation~(\ref{fractionEst}) then 
represents a tremendous suppression with increasing $N$ of vacua
with nucleation rates low enough to maintain metastability.

Note that we have ignored gravitational effects
in our analysis of tunneling rates.  With low barriers and rapid
bubble nucleation this is generally a good approximation.  It should
be noted, though, that in a de Sitter vacuum with a relatively flat
barrier the Hawking-Moss bounce provides an alternative mode of vacuum
decay.  The corresponding decay exponent is
\begin{equation}
  B_{\rm HM} = {3\over 8 G_N^2} \,{V_{\rm barr}
       - V_{\rm fv} \over V_{\rm barr} V_{\rm fv}}  \, ,
\end{equation}
where $V_{\rm fv}$ and $V_{\rm barr}$ are the values of the potential
in the false vacuum and at the top of the barrier, while $G_N$ is
Newton's constant.  Our numerical studies indicate that the barrier
height relative to the false vacuum (i.e., $V_{\rm barr} - V_{\rm
  fv}$) has a power law falloff  with $N$ quite similar to that found
in the effective wall tension $s$, thus implying a similar enhancement
of the Hawking-Moss transition rate.

\section{Possible Implications for the String Landscape}
\label{implications}

The results we have described so far are a general feature of quantum
field theories.  Our motivation for studying random multi-field
potentials, however, is the landscape of string theory. In this
section, therefore, we assess the applicability and implications of
our findings for the string landscape.

Not only does the enormous number of local minima in the high
dimensional string theory landscape (which can serve as the endpoints
of tunneling events) offer the possibility of significant
enhancement of tunneling probabilities,\footnote{The density of vacua
  near the conifold point, $\rho_{\rm conifold}$, in a one-dimensional
  moduli space is described by $1/[r^2(C+\log r)^2]$, where $r$ is the
  distance from the conifold point \cite{Denef:2004ze}; applying this
  result near a generic point along the conifold locus in an
  $n$-dimensional moduli space shows the rapid growth in the number
  of vacua, $\int d^nr \,\rho_{\rm conifold}\,$, with $n$.}  but we have
also seen that such large numbers of fields result in the exponential
suppression of tunneling barriers. The question we now briefly
consider is the extent to which the random potentials we have analyzed
in the previous sections provide accurate insight into properties of
the string landscape.

For definiteness, focus on a standard flux compactification of type
IIB string theory on a Calabi-Yau manifold $M$.  With $G = F - \tau
H$, $F$ and $H$ the RR and NS-NS three-form fluxes, and $\tau$ the
axio-dilaton, the Gukov-Vafa-Witten superpotential $W$ is given
by~\cite{Gukov:1999ya}
\begin{equation}
W = \int G \wedge \Omega  \, ,
\label{superpot}
\end{equation}
where $\Omega$ is the holomorphic 3-form on $M$. The associated flux
potential $V$ is given by
\begin{equation}
V_{M} = e^{K}(D_{\rho} W D^{\rho} {\overline W} - 3 |W|^2)  \, .
\label{susyV}
\end{equation}

Assume that $G$, together with additional contributions (e.g.~D3-brane
instantons or wrapped D7-branes), stabilizes all moduli
\cite{Denef:2004dm}. Then consider $\{V_M\}$ as $M$ varies among all
Calabi-Yau's, $G$ varies over flux values, the K\"ahler stabilizing
contributions vary over all possibilities, and supersymmetry-breaking
effects similarly scan a broad range.
 With all such variations in play, the local minima of
$\{V_{M}\}$ will sweep through a large class of vacua. In
the vicinity of any such vacuum state, we can expand the effective
potential, yielding a field theory model of the form
(\ref{Vexpansion}). Now, considering all such expansions that arise
from all local minima of the collection $\{V_{M}\}$, we expect the
expansion coefficients to randomly vary. Indeed, such random variation
inspired the ansatz we chose in our numerical studies. But in applying
the results of the previous sections to the string landscape, there
are a number of details and complications that deserve further
attention.

First, we have assumed that the action for tunneling trajectories is
well approximated by our proxy: twice that of a straight line in field
space connecting the minima being studied and the optimal saddle point
on the surrounding barrier. Yet, tunneling trajectories in
multidimensional field spaces are notoriously subtle and can exhibit
unexpected features; explicit examples in the landscape are the
conifunneling trajectories between monodromy-related flux vacua found
in \cite{Ahlqvist:2010ki} (see also \cite{ Danielsson:2006xw,
  Johnson:2008kc}). An additional, and potentially pivotal,
complication is that the different flux vacua that are not monodromy
related are generally minima of distinct potentials. Physically, such
transitions invoke features not captured by our local field theoretic
model, including for example the nucleation of branes to absorb
changes in flux \cite{Brown:1988kg,de Alwis:2006cb}. These effects
might significantly affect the tunneling action, and possibly mitigate
the field theory instabilities we have identified.

Second, we have assumed that as the stabilizing contributions to a
given model are varied the effective potentials around local minima
will have expansions that are well modeled by random polynomials. Is this
correct? To be concrete, consider the part of the potential arising
from the flux, $G$. On a given Calabi-Yau, $M$, there are dim $H^{3}(M)$ fluxes
that enter $G$. As those fluxes vary, we have far fewer free
parameters than the coefficients in Eq.~(\ref{Vexpansion}). However,
the associated minima of $V$ will occur at different locations, $p$,
in the moduli space of $M$. As the holomorphic form $\Omega$ depends
on $p$, the coefficients in a local expansion of the superpotential,
Eq.~(\ref{superpot}), and the corresponding potential,
Eq.~(\ref{susyV}), will also vary with $p$. So, the local potential
will be randomized both by the varying fluxes and by the varying
values of the period vector over the moduli space. It seems reasonable
to us that this will result in local expansions well modeled by the
random potentials invoked in Sec.~\ref{numerics}, but we do not have a
firm argument. We will return to this issue in a forthcoming work,
where we will explicitly check this for specific
Calabi-Yau's with low-dimensional moduli spaces.

Third, we have taken a canonical form for the kinetic terms
in our field theoretic models. It is well known, however, that string
vacua are densest in the vicinity of the conifold locus, where the
classical moduli space metric suffers from a curvature singularity. In
particular, near a generic point on the conifold locus we can
choose local coordinates $(Z^1, Z^2, \dots ,Z^P)$ on the moduli space
such that $Z^1 = 0$ labels the conifold. Near $Z^1 = 0,$ the moduli space
metric $G$ behaves as:

\begin{equation}
G^{11}(Z) \sim {\rm ln} (|Z^1|^2)  \, .
\end{equation}
The local form of the action then takes the form
\begin{equation}
\int \sqrt {-g} [g^{\mu \nu} G_{i j} \,\partial_\mu Z^i \,\partial_{\nu} Z^j - V(Z)],
\end{equation}
where $g$ is the space-time metric. In this expression the
coordinates $Z$ are the moduli space representation of the scalar
fields $\phi$ and $V$ is their flux potential.  At any non-singular
point we can, of course, use a local change of field variables to
absorb $G$ into the $Z$, yielding a canonical kinetic term. But as we
approach a conifold point, this change of variables corresponds to
reducing the barrier heights in $V$ (assuming $V$ is continuous and is
being expanded about a local minimum) and thus increases tunneling
rates. In the regions of moduli space that are most densely populated
with string vacua, we therefore expect the non-canonical kinetic terms
to augment the destabilization we have found.

Fourth, since the flux potential $V$ is derived from the
superpotential $W$, and it is $W$ that directly incorporates flux
values, a more accurate representation of the landscape arises from
randomly varying coefficients in a local expansion of $W$, and then
using the result to calculate the random potentials. In principle, the
relationships between the coefficients in $V$, which reflect its
origin in $W$, could alter our findings.

We have undertaken such an analysis. Specifically, we considered a
theory with $N$ chiral superfields $\Phi_j$.  The superpotential $W$
was taken to be a polynomial
\begin{equation}
    W = \frac12\sum_i  C_{ii}^{(2)} \Phi_i^2 
      + \frac13 \sum_{ijk} C_{ijk}^{(3)} \Phi_i \Phi_j \Phi_k 
      + \frac14\sum_{ijkl} C_{ijkl}^{(4)} \Phi_i \Phi_j \Phi_k \Phi_l
     + ... 
\label{Wexpansion}
\end{equation}
The $C_{ii}^{(2)}$ were randomly chosen real numbers in the range
$[0,1]$, while $C_{ijk}^{(3)}$ and $C_{ijkl}^{(4)}$ were complex numbers
whose real and imaginary parts were taken randomly from the interval
$[-1,1]$.  The scalar field potential derived from $W$ was then
truncated at quartic order and analyzed in a fashion similar to our
non-supersymmetric potentials.  With up to $N=5$ superfields we again
find a power law falloff with $N$.  Because the $N$ chiral superfields
correspond to $2N$ real scalar fields, instead of writing the fits to
the data as in Eq.~(\ref{s-fit}) we write, e.g.,
\begin{equation}
     s_{\rm median} = C_{\rm tension} (2N)^{-\alpha} \, . 
\end{equation}
As can be seen from the data in our tables, the various fit parameters
are rather similar to those for the nonsupersymmetric case, with the
most notable difference being that $s_{\rm median}$ varies as
$N^{-3.16}$, compared to $N^{-2.66}$ in the non-supersymmetric
case\footnote{Because we have not included any supersymmetry-breaking
  terms in the potential, the vacua here are actually stable, and our
  data correspond to domain walls rather than bubble walls.  However,
  we do not expect the picture to be changed materially when the vacua
  are lifted.}.

This example also illustrates the challenges of investigating random
potentials for large numbers of fields. In all of our studies,
supersymmetric or not, computational considerations have forced
us to only probe a limited range of values of $N$, the number of fields,
and work with truncated potentials. We are assuming that the pattern
we have found, as evidenced in
Figs.~\ref{QuarticTensionAndHeight}-\ref{QuarticDistributions}, will
continue to hold as these constraints are relaxed.

\section{A multiverse explanation of the cosmological constant?}
\label{Multiverse}

Our results have immediate implications for attempts to find a
multiverse solution to the cosmological constant problem in such field
theory models.  The argument for such a solution is that even if the
natural scale for the cosmological constant in a typical vacuum is
Planckian, one might expect to find vacua with $\Lambda \sim
10^{-120}$ in Planck units if the number of vacua is much greater than
$10^{120}$.  Because we want a vacuum that has not only a small
$\Lambda$ but also a long lifetime, what we actually need is that the
number of truly metastable vacua be much greater than $10^{120}$.

Let us suppose that the number of vacua is ${\cal N}_{\rm vac} \sim
{\cal F}^N$, with perhaps ${\cal F}=10$. For metastability, we require that the 
tunneling exponent $B$ be no smaller than a value $B_{\rm min}$ of 
order unity.  The number of metastable vacua is then 
\begin{equation}
    {\cal N}_{\rm vac} \, f(B_{\rm min}) 
    \sim {\cal F}^N \, e^{-\beta B_{\rm min}N^\alpha}
\end{equation}
with $f(B)$ given by Eq.~(\ref{fractionEst}).  The requirement that this
be greater than $10^{120}$ can be written as 
\begin{equation}
     N - {b \over \ln {\cal F}} N^\alpha > 120 \left({\ln 10 \over \ln {\cal F}} \right)
\end{equation}
with 
\begin{equation}
   b = \beta B_{\rm min} = \left( 10^{-3} \gamma \over C_{\rm tension} \right)
               \lambda B_{\rm min} \sim  10^{-3} \lambda \, .
\end{equation}

\begin{figure}
   \centering
   \includegraphics[width=5.5in]{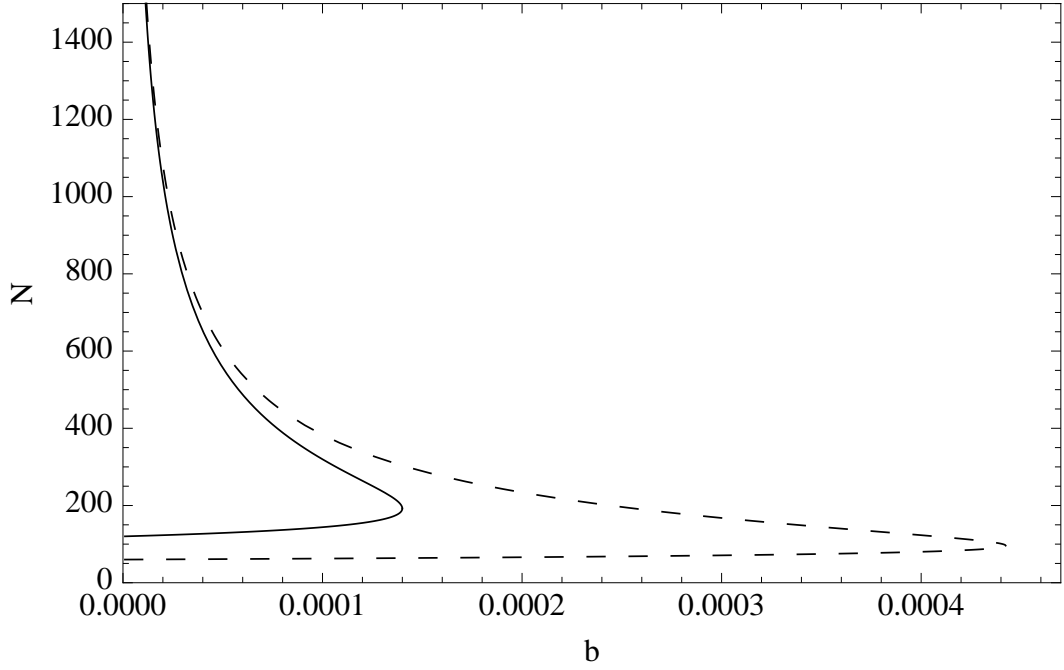}
   \caption{Parameter ranges allowing multiverse explanations of the
cosmological constant.  With $\alpha=2.66$ and ${\cal F}=10$, a sufficient number
of metastable vacua is only possible for parameters in the region to the 
left of the solid line.  This region is extended to the dashed line if
${\cal F}=100$. }
   \label{QuarticRange}
\end{figure}

This clearly needs large $N$, at least more than 120 if ${\cal F}=10$
as in the usual analysis.  The new feature here is that, because of
the enhancement of the tunneling rate at large $N$, taking $N$ to be
too large makes matters worse rather than better.  In fact, there may
be no value of $N$ for which this condition is satisfied.  This is
illustrated in Fig.~\ref{QuarticRange}, where we have taken
$\alpha=2.66$, the power we obtained from the analysis of quartic
non-supersymmetric potentials.  For ${\cal F}=10$ the allowed values
of $N$ and $b$ correspond to the region to the left of the solid
curve.  We see that there are no acceptable values of $N$ for $b > 1.4
\times 10^{-3}$, and that the range of allowed $N$ is relatively
restricted until $b$ falls well below this value.  If instead we take
${\cal F}=100$, the allowed region extends to the dashed line, with
the maximum allowed $b$ increased by roughly a factor of four.

\begin{figure}
   \centering
   \includegraphics[width=5.5in]{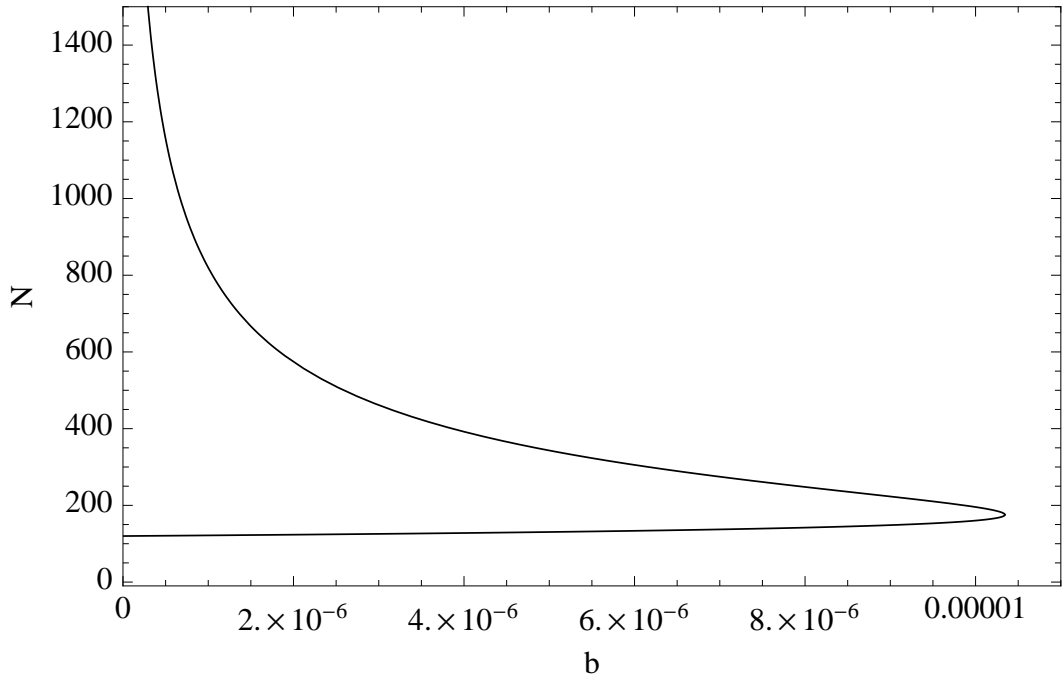}
   \caption{Like Fig.~\ref{QuarticRange}, but for $\alpha=3.16$ and 
${\cal F} =10$.}
   \label{SUSYRange}
\end{figure}

The effect of varying the power $\alpha$ can be seen in Fig.~\ref{SUSYRange},
where we have set $\alpha = 3.16$, the value from our supersymmetric 
data.   Although the curve is similar to that in the previous case, the 
values of $b$ have fallen by more than an order of magnitude.

\section{Conclusions}
\label{Conclusions}

Motivated by the string landscape, we have undertaken a study of the
stability of vacua in multi-field quantum theories. Our
study has focused on random potentials with the number of fields $N
\le 10$ and on polynomial expansions that include no more than quartic
field contributions. Even with these constraints, coming from
computational feasibility, the data we have accumulated provide
evidence that transition rates are so rapidly enhanced as a function
of $N$ that all but an exponentially small fraction of generic local
minima in such quantum field theories are unstable to rapid decay.

One consequence is that the range of parameters that give a sufficient
number of metastable vacua to provide a natural solution to the
cosmological constant problem is severely restricted.  For example, if
the moduli space dimension $N=500$ and there are ${\cal F}^N=10^{500}$
vacua, the ``coupling'' $\lambda$ characterizing the landscape
must be less than 1/20 or so; with $N = 2000$, $\lambda$ must be an
order of magnitude smaller.  Alternatively, if $\lambda=1/2$ and
$N=500$, then ${\cal F}$ must be greater than $6\times 10^6$.

These considerations are potentially relevant to any model invoking
anthropic explanations for the cosmological constant in which the
required diversity of vacua is due to the model containing a high
dimensional moduli space. However, since our results have been
obtained in the context of a field theory with multiple scalar fields,
the impact of similar considerations on the ability of the string
landscape to offer a natural solution to the cosmological constant
puzzle will require further study.

\begin{acknowledgments}
We thank Pontus Ahlqvist, Robert Brandenberger, Adam Brown, Alex Chen,
Frederik Denef, Zhihua Dong, Kurt Hinterbichler, Vojkan Jaksic, Dmitry
Jakobson, Luchang Jin, Dan Kabat, Zhongjie Lin, Bob Mawhinney,
Massimo Porrati, I-Sheng Yang, Hantao Yin, Jianglei Yu,
and Daiqian Zhang for helpful discussions and comments. This work was
supported in part by the U.S. Department of Energy under grants
DE-FG02-85ER40237 and DE-FG02-92ER40699.  Parts of the computation
were carried out on Fermilab LQCD clusters.
\end{acknowledgments}

\end{document}